\documentclass[11pt]{article}
\usepackage{moriond,epsfig}

\def\Journal#1#2#3#4{{#1} {\bf #2}, #3 (#4)}

\def\NPB{{\em Nucl. Phys.} B}
\def\PLB{{\em Phys. Lett.}  B}
\def\PRL{\em Phys. Rev. Lett.}
\def\PRD{{\em Phys. Rev.} D}

\def\JHEP{\em JHEP}
\def\be{\begin{equation}}
\def\ee{\end{equation}}
\def\bea{\begin{eqnarray}}
\def\eea{\end{eqnarray}}

\begin{document}
\vspace*{4cm}
\title{Latest Jets Results from the Tevatron}

\author{ Christina Mesropian \\
   on behalf of the CDF and D0 Collaborations}

\address{The Rockefeller University,1230 York Avenue,\\
New York, NY, 10065, USA}

\maketitle\abstracts{A comprehensive overview of the latest aspects of  jet physics
  in proton-antiproton collisions at $\sqrt{s}=$1.96 TeV is presented. In particular,
 measurements of the inclusive jet production, dijet and multi-jet production, and jet substructure studies are discussed.}

\section{Inclusive Jet Production}\label{subsec:prod}
The experimental measurements of jet cross section at the Tevatron provide stringent test of QCD predictions, 
information on the strong coupling constant, $\alpha_S$, and constraints on proton parton distribution functions, PDFs.
The inclusive jet cross section measurements were performed by the CDF collaboration~\cite{cdf-incljet}$^,$~\cite{cdf-incljet-kt} using midpoint cone~\cite{midpoint} and
 $k_T$~\cite{kt}  algorithms and by D0 collaboration using  the midpoint algorithm~\cite{d0-incljet}. Both experiments extended measurements to the forward rapidity regions. 
 The systematic uncertainties in these measurements are dominated  by  the uncertainty 
in the jet energy scale. The extensive efforts to determine jet energy scale, using single particle response technique in the case of CDF, and $\gamma +jet$
event calibration method at D0, allowed to  achieve the jet energy scale uncertainties
 of 2-3\% and 1-2\%, respectively.  The understanding gained by these measurements is important for any analyses which have jets as an object of interest.
%--------------------------------------------------
\begin{figure}[htbp]
\includegraphics[width=0.35\columnwidth]{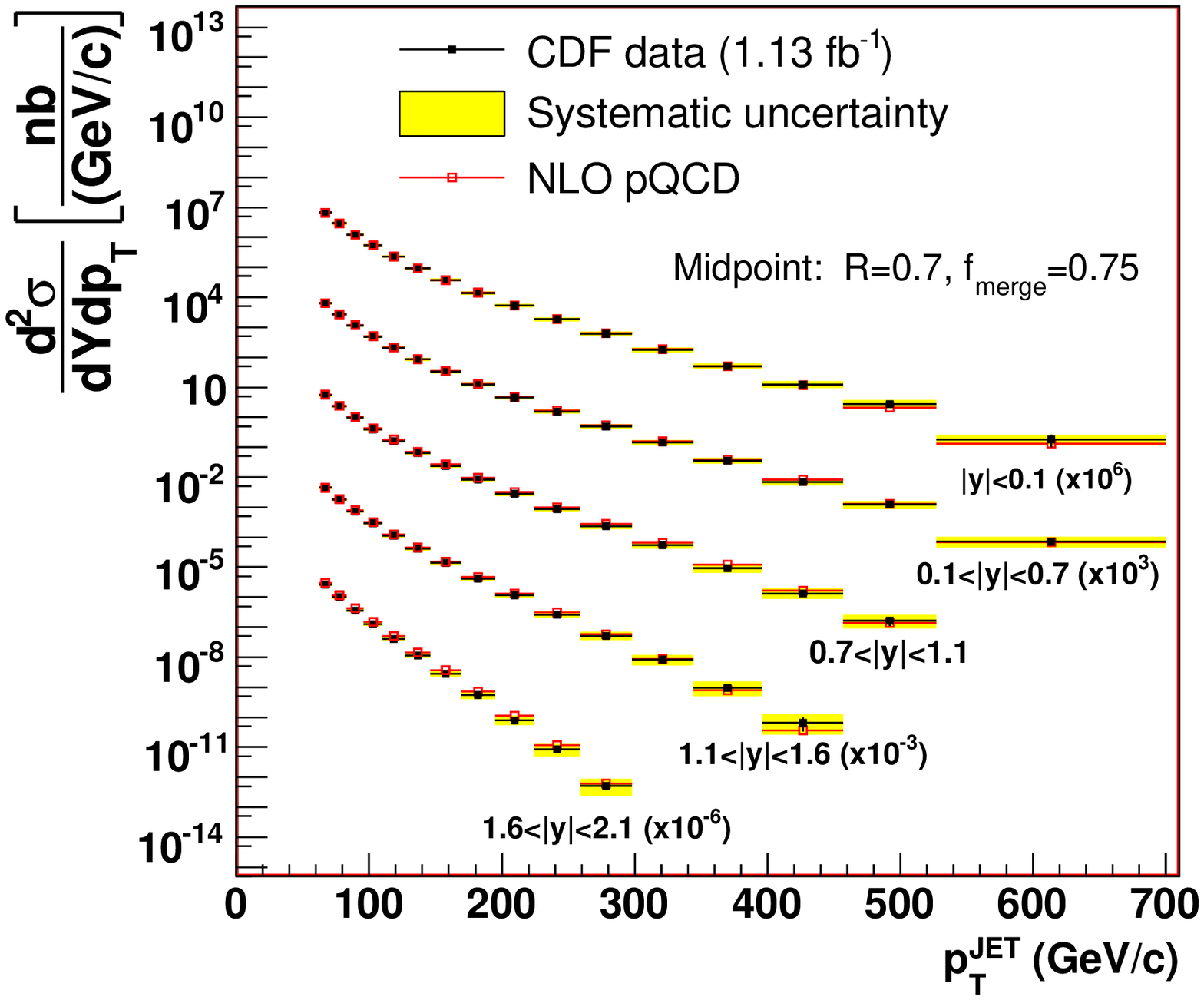}
\includegraphics[width=0.60\columnwidth]{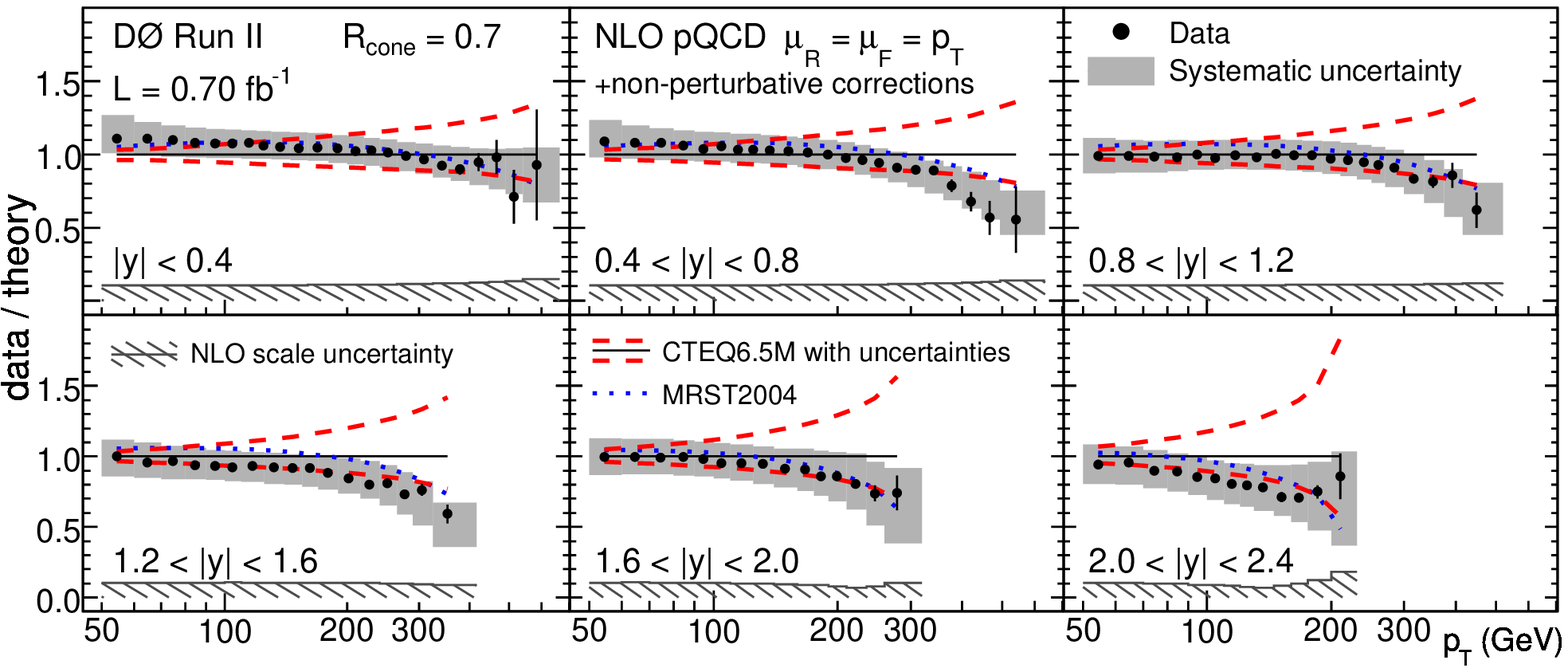}
\caption{Measured inclusive jet differential cross sections in five rapidity regions by CDF compared to NLO QCD predictions (left); Ratios of the measured cross sections over NLO QCD predictions by D0 (right).}\label{mesropian_christina.fig1}
\end{figure}
%--------------------------------------------------

Fig.~\ref{mesropian_christina.fig1} shows a comparison of the measured cross sections to the theoretical predictions. 
The measurements are found to
 be in agreement with NLO QCD predictions for both experiments and for different clustering algorithms.
The experimental uncertainties are lower than the  uncertainties associated with the theoretical predictions.
 Since inclusive jet measurements allow to constrain PDFs of the proton, 
especially gluon densities at high $x$, ($x\geq$0.25),
two groups performing  global QCD analyses to determine PDFs included these Tevatron
 measurements in their compilation, with resulting PDFs referred as MSTW2008 and CT09. Inclusion of the Tevatron measurements
 lead to somewhat softer high-$x$ gluons than the ones previously available. 

The inclusive jet cross section is directly related to the measurement of the strong coupling constant.
The CDF collaboration performed this analysis using the 1994-95 data, and D0 recently published a new $\alpha_S$ 
determination~\cite{d0-alpha-S} based on the inclusive jet cross section measurement discussed above. The value of the strong 
coupling constant is determined from sets of inclusive jet cross section data points by minimizing the $\chi^2$ 
function between data and the theoretical results. In order to avoid the complications arising from the $\alpha_S$ 
dependence on PDF determinations, only 22 data points out of 110 
were kept for $\alpha_S$ determination.
 This measurement provides the most precise result for the strong coupling constant from the hadron 
colliders $\alpha_S(M_Z) =0.1161\pm^{0.0048}_{0.0041}$.

The CDF~\cite{cdf-dijet} and D0~\cite{d0-dijet} experiments used the dijet invariant mass distribution to search for resonances decaying into jets. In the case of D0, measurements of the dijet angular distributions are performed in different regions of the dijet invariant mass. 
A good agreement between data and theory, which translates into improved limits in different models,  is observed for both experiments.

\section{Multi-Jet Production}

We present a measurement  by the D0 collaboration of the differential inclusive three-jet cross section as a function of the invariant three-jet mass ($M_{3jet}$). 
The data set corresponds to an integrated luminosity of 0.7 fb$^{-1}$.
The measurement is performed in three rapidity regions ($\mid{y}\mid<$0.8, $\mid{y}\mid<$1.6, and $\mid{y}\mid<$2.4) 
and in three regions of the third jet transverse momenta.
The events are required to have $p_T$ of the leading jet  larger than 150 GeV and for any pair of jets to be well separated in $y-\phi$ space.  The comparison of the experimental data with the NLO theoretical predictions shows a reasonable agreement, see Fig.~\ref{mesropian_christina.fig1}. 

%--------------------------------------------------
\begin{figure}[htbp]
\centerline{
\includegraphics[width=0.4\columnwidth]{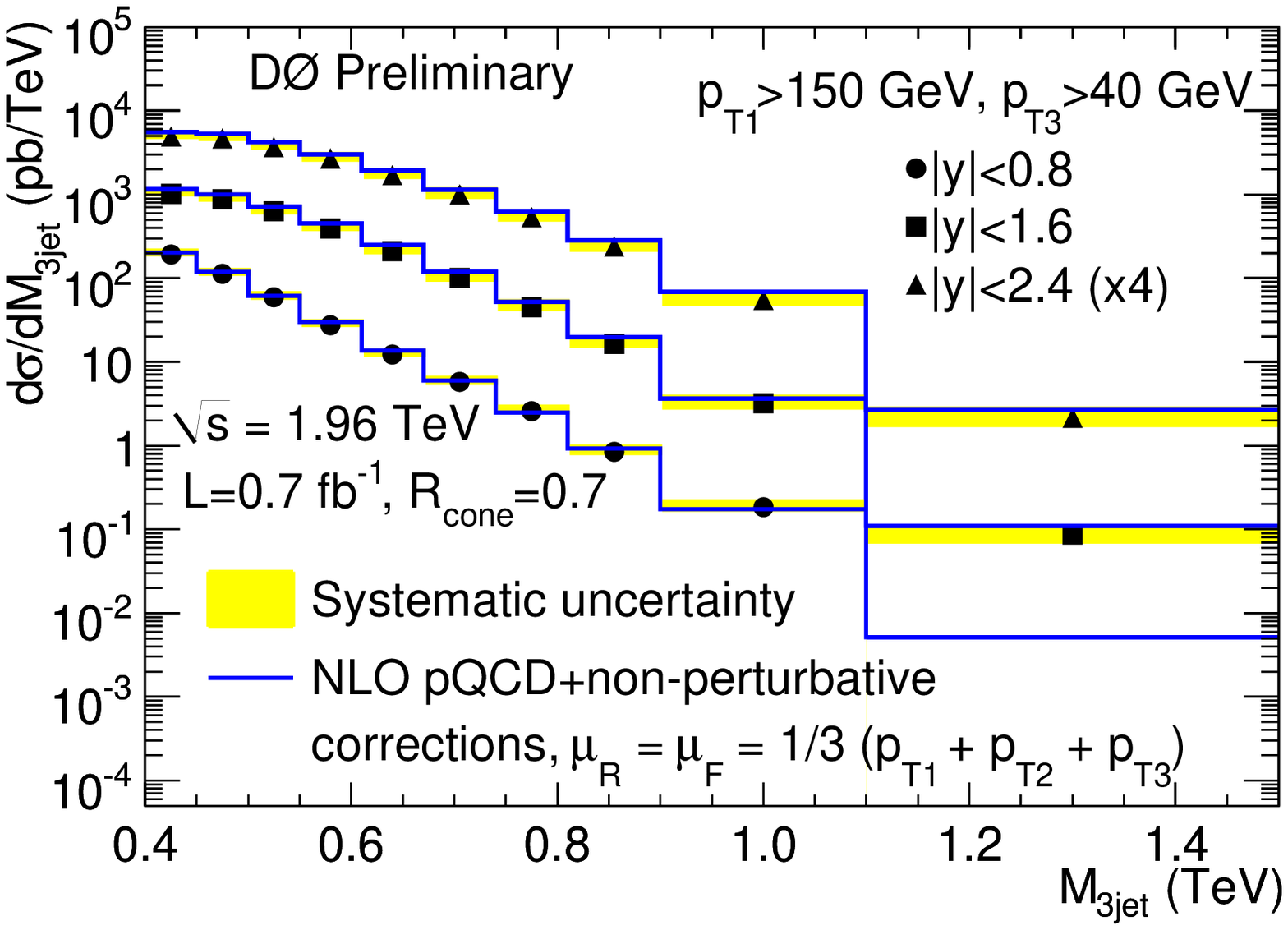}
\includegraphics[width=0.4\columnwidth]{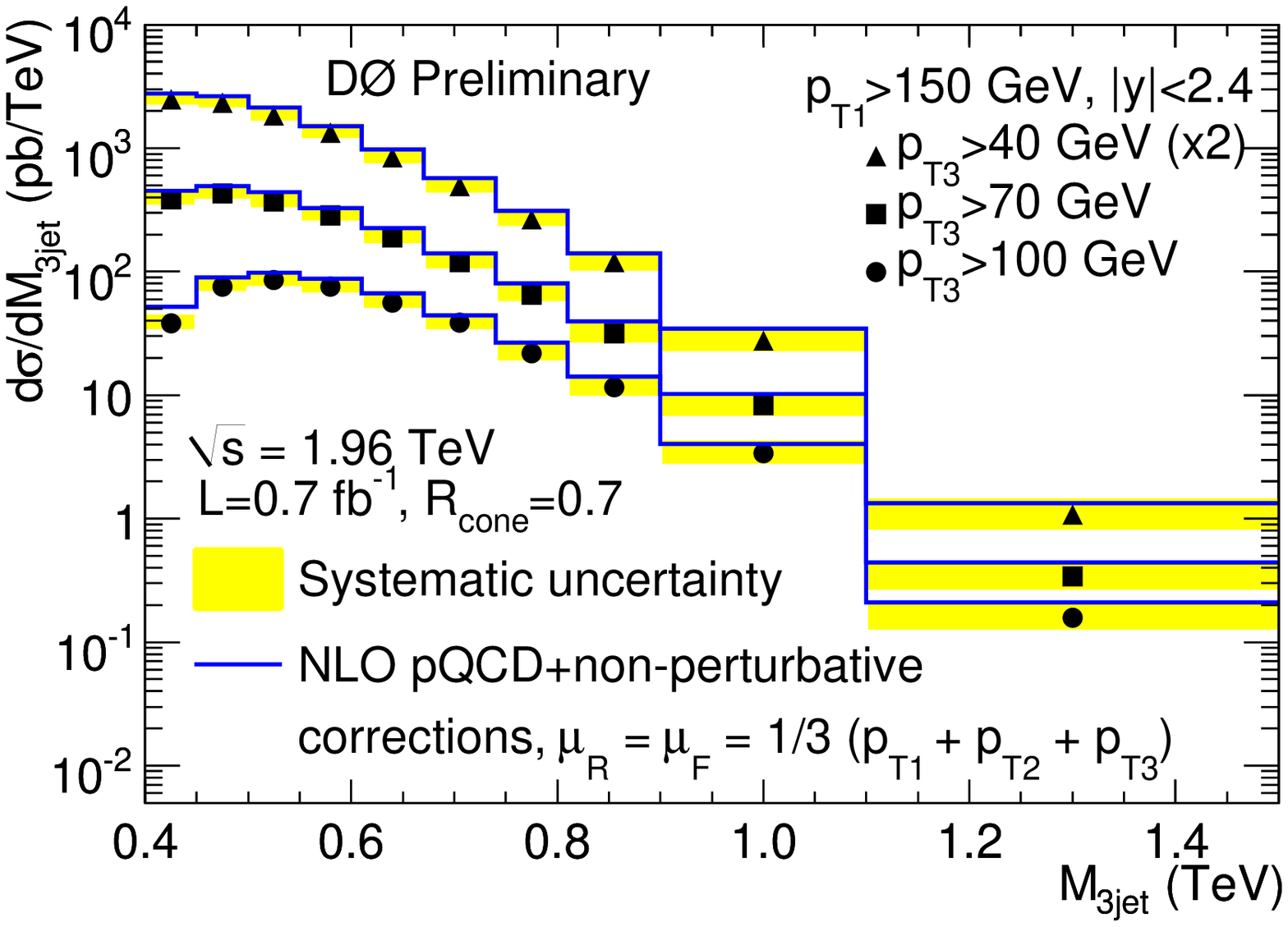}
}
\caption{Three-jet mass cross section in regions of jet rapidities (left), and third jet $p_T$ (right). 
Full lines correspond to the NLO calculations with NLOJET++ and MSTW2008 PDF set.}\label{mesropian_christina.fig1}
\end{figure}
%--------------------------------------------------

\begin{figure}[htbp]
\includegraphics[width=0.85\columnwidth]{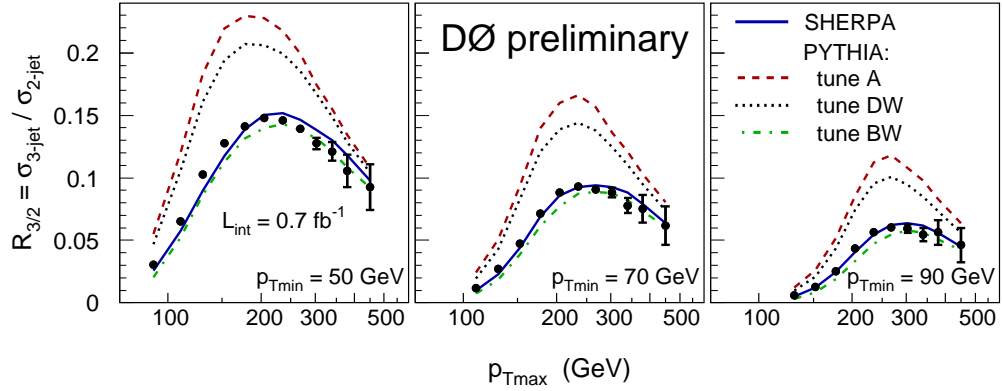}
\caption{$R_{3/2}$ ratio measured as a function of the leading jet $p_{Tmax}$ for different
 $p_{Tmin}$ requirements for the other jets. Predictions from SHERPA and PYTHIA 
(three tunes using the virtuality-ordered parton shower)
 are compared to the data.}\label{mesropian_christina.fig2}
\end{figure}

 Using the same data sample 
the D0 collaboration also performed a measurement of the ratios of the multi-jet cross sections. 
The inclusive n-jet event sample (for n = 2,3) is  defined by all events with n or more jets 
with $p_T> p_{Tmin}$ and $\mid y\mid<$2.4. The rapidity requirement restricts the jet phase space 
to the region where jets are well-reconstructed in the D0 detector and the energy calibration is known to 1.2 - 2.5\%.
The ratio of cross sections is less sensitive to experimental and theoretical uncertainties than the individual cross sections, 
due to cancellations of correlated uncertainties. $R_{3/2}$ is measured as a function of the leading jet $p_T$ in an event, 
$p_{T_{max}}$. Since the variable  $p_{T_{max}}$ is independent of 
the jet multiplicity, all events which belong to a given $p_{T_{max}}$ bin for the inclusive trijet event 
sample also belong to the same $p_{T_{max}}$ bin for
the inclusive dijet event sample. Given the definitions above for inclusive n-jet event samples, $R_{3/2(p_{Tmax})}$
 equals the conditional probability for an inclusive dijet event (at $p_{Tmax}$) to contain a third jet with $p_T > p_{Tmin}$. 
The data is well described by the SHERPA event generator (using default settings) with tree-level matrix elements 
for 2-, 3-, and 4-jet production. For the PYTHIA event generator, the results depend strongly on the chosen parameter 
tune. Commonly used tunes (for both the angular-ordered and the $p_T$ -ordered parton shower), see Fig.~\ref{mesropian_christina.fig2}, overshoot the measured 
ratios significantly over the whole $p_{Tmax}$ range for all $p_{Tmin}$ requirements.
 %--------------------------------------------------
%--------------------------------------------------

\section{Study of Substructure of High $p_T$ Jets}

The study of high transverse momentum ($p_T$) massive jets provides an important test of pQCD and gives insight
 into the parton showering mechanism. In addition,
 massive boosted jets compose an important background in searches for various new physics 
models, the Higgs boson, and highly boosted top quark pair production. 
%--------------------------------------------------
\begin{figure}[htbp]
\centerline{\includegraphics[width=0.5\columnwidth]{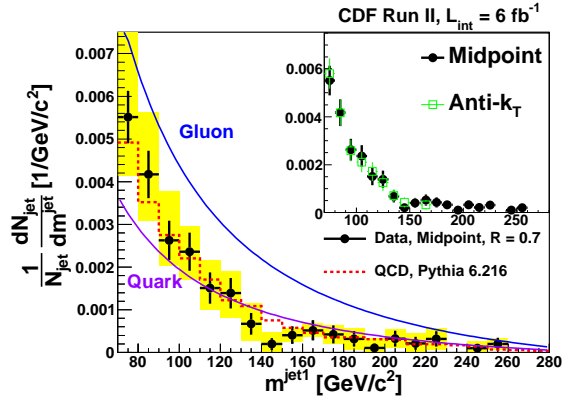}}
\caption{The normalized jet mass distribution for midpoint jets with $p_T>$ 400 GeV/c.
The theory predictions for the jet functions for quarks and gluons are shown as solid curves 
and have an estimated uncertainty of ~30\%. The inset compares  midpoint and $anti-k_t$ jets.}
\label{mesropian_christina.fig4}
\end{figure}
%--------------------------------------------------
Particularly
relevant is the case where the decay of a heavy resonance produces high-$p_T$ top quarks that
decay hadronically. In all these cases, the hadronic decay products can be detected as a single jet
 with substructure that differs from pQCD jets once the jet $p_T$ is greater than 400-500 GeV/c.
The CDF collaboration performed measurement of substructure of jets with $p_T>$400 GeV/c by studing 
distributions of the jet mass
and measuring angularity, the variable describing the energy distribution inside the jet, and planar 
flow, the variable  differentiating between two-prong and three-prong decays.  
At small cone sizes and large jet mass, these variables are expected to be quite robust against soft radiation
 and allow, in principle, a comparison with theoretical predictions in addition to comparison with MC results.
Jets are reconstructed with the midpoint cone algorithm (cone radii R=0.4, 0.7, and 1.0) and 
with the $anti-k_t$ algorithm~\cite{anti-kt} (with distance parameter R=0.7).
Events are selected in a sample with 6 fb$^{-1}$ based on the inclusive jet trigger. There is a good agreement
 between the measured $m^{jet_1}$ distribution with the analytic predictions for the jet function and with PYTHIA MC predictions.
 The midpoint and $anti-k_t$ algorithms have very similar jet substructure distributions for high mass jets, 
see Fig.~\ref{mesropian_christina.fig4}. The angularity distribution shown on Fig.~\ref{mesropian_christina.fig5}(left) in addition 
to reasonable agreement data and PYTHIA MC also demonstrates that the high mass jets coming from light quark 
and gluon production are consistent with two-body final states and that further rejection against high mass 
QCD jets can be obtained by using the planar flow variable, Fig.~\ref{mesropian_christina.fig5}(right).
%--------------------------------------------------
\begin{figure}[htbp]
\centerline{\includegraphics[width=0.45\columnwidth]{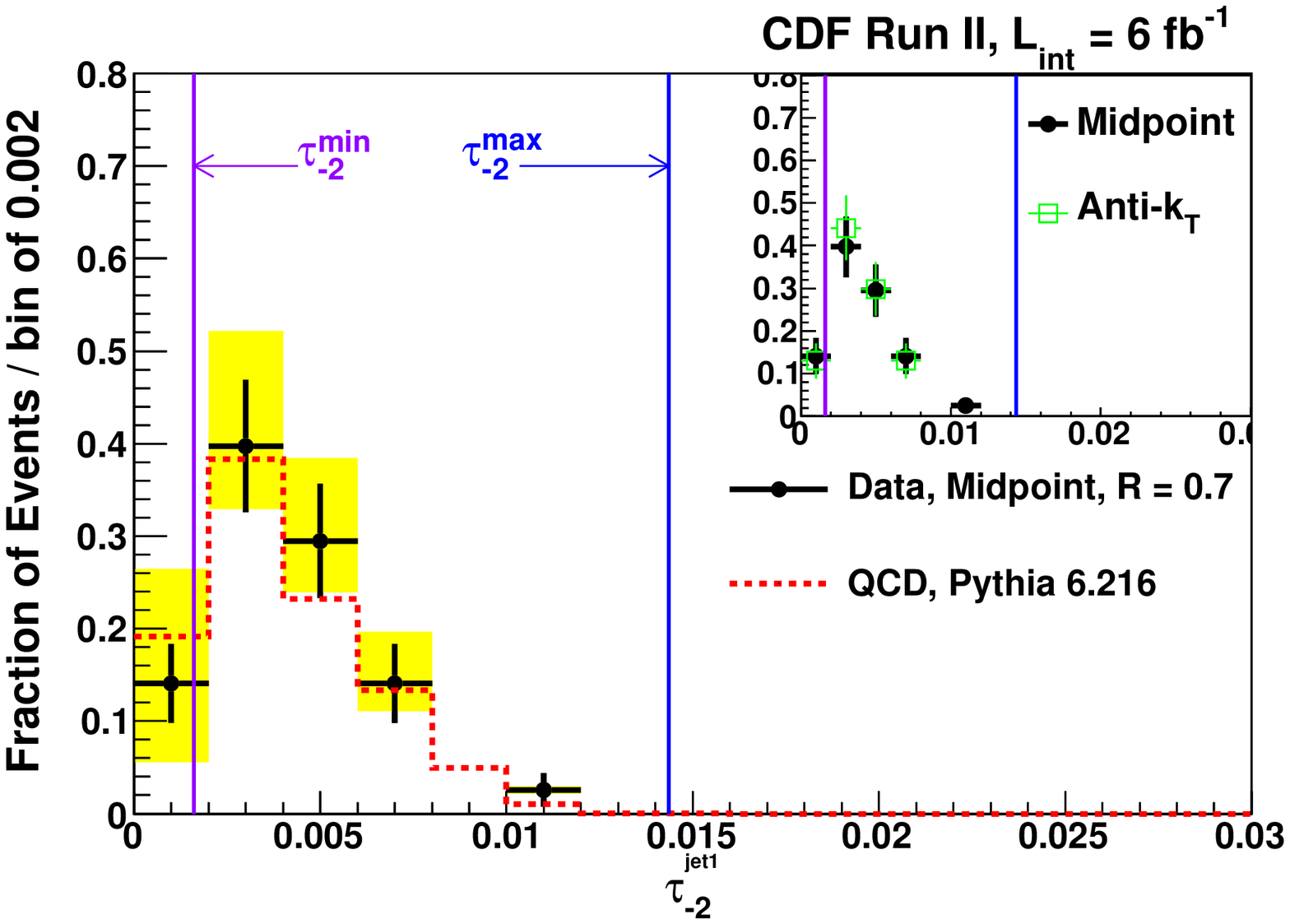}
\includegraphics[width=0.45\columnwidth]{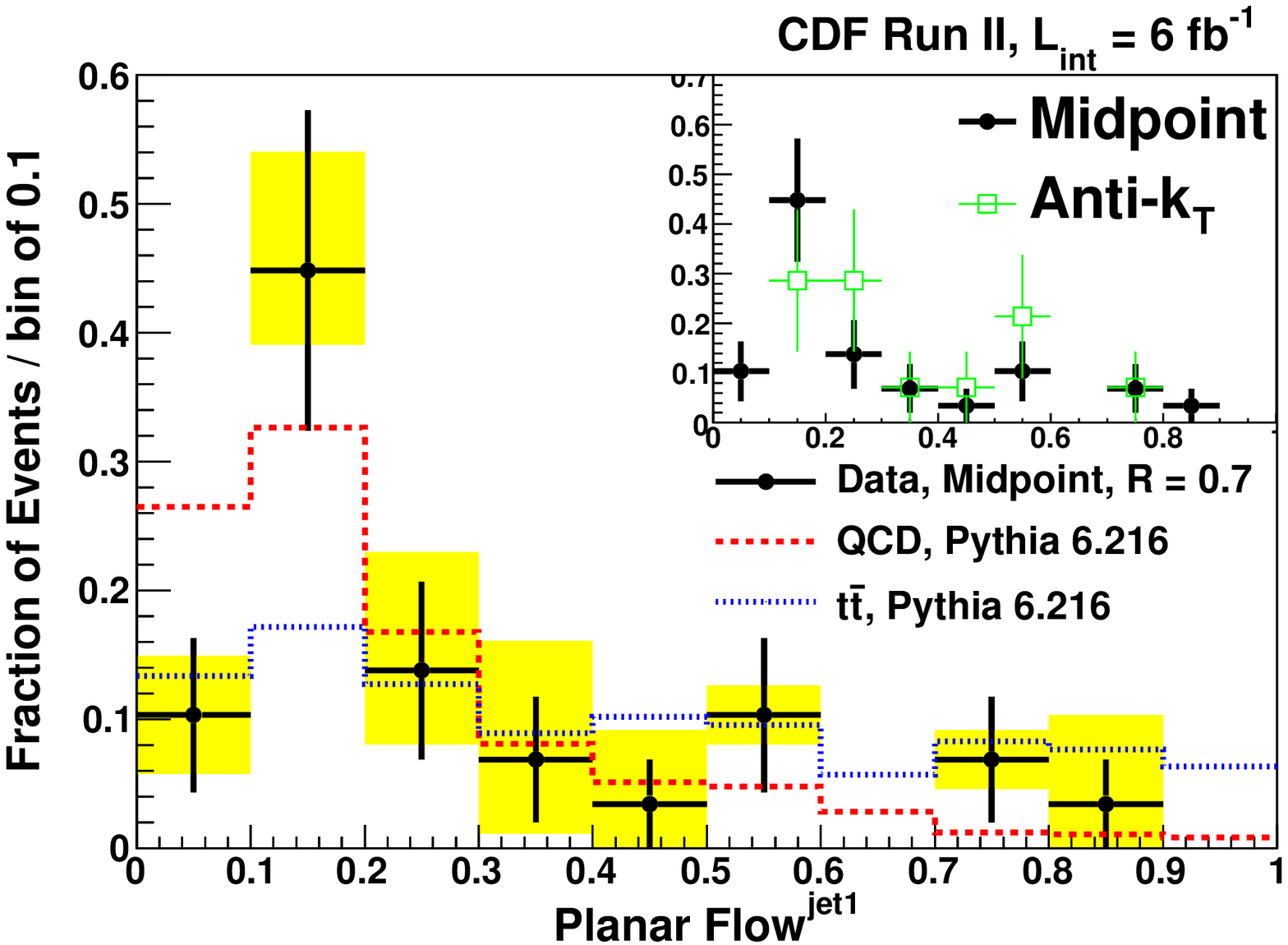}
}
\caption{The angularity distribution for midpoint jets with $p_T>$400 GeV\/c. The $t\bar{t}$ rejection cuts and requirement 
for 90 GeV/c$^2<m^{jet1}<$120 GeV/c$^2$ are applied. The PYTHIA calculation (red dashed line) and the pQCD kinematic endpoints are shown (left); 
The planar flow distributions after applying the top rejection cuts and requiring 130 GeV/c$^2<m^{jet1}<$210 GeV/c$^2$. 
PYTHIA QCD (red dashed line) and $t\bar{t}$ (blue dotted line) jets are shown (right).}\label{mesropian_christina.fig5}
\end{figure}
%--------------------------------------------------

\section*{Acknowledgments}
I would like to thank organizers for their kind invitation to the conference.

\end{document}